\documentclass[a4paper,11pt]{article}
\usepackage{pos}
\graphicspath{{figures/}}
\usepackage{caption}
\usepackage{subcaption}
\usepackage{cleveref}

\title{Measurement of $b \to u \ell \nu_{\ell}$ at Belle II}

\author*{Andrea Fodor \footnote[2]{On behalf of the Belle II Collaboration.}}

\affiliation{McGill University,\\
  845 Sherbrooke Street West, Montreal, Canada}

\emailAdd{afodor@physics.mcgill.ca}

\abstract{The semileptonic $B$ meson decays of the type $B \to X_u \ell \nu$ are important for measuring the magnitude of the CKM matrix element $V_{ub}$. Recent measurements of inclusive $B \to X_u e \nu_e$ and hadronically tagged $B^0 \to \pi^- \ell^+ \nu$ from the Belle II experiment are presented. The results are obtained using the data collected at Belle II between March 2019 and March 2020, with a total integrated luminosity of 37.8~fb$^{-1}$. These preliminary measurements are used to validate the detector and reconstruction software performance. The measured preliminary branching fraction of $B^0 \to \pi^- \ell^+ \nu$ decays is $(1.58 \pm 0.43_{\textrm{stat}} \pm 0.07_{\textrm{sys}})\cdot 10^{-4}$, which is in agreement with the world average, $(1.50 \pm 0.06)\cdot 10^{-4}$. }

\FullConference{%
  BEAUTY2020\\
  21-24 September 2020\\
  Kashiwa, Japan (online)}

\begin{document}
\maketitle

\section{Introduction}
Belle II  is a $B$-factory experiment located at the KEK laboratory in Tsukuba, Japan. The SuperKEKB accelerator collides electrons and positrons at a center of mass (COM) energy of $10.58$~GeV\footnote{Natural units, with $c = 1$ are used throughout this report.} at the interaction point inside the Belle II detector. At this COM energy the $\Upsilon(4S)$ is copiously produced. The $\Upsilon(4S)$ meson subsequently decays almost exclusively to $B$ meson pairs, $B^+B^-$ or $B^0\bar{B}^0$. Belle II started its major physics run in March 2019. The results presented here use the data collected between March 2019 and March 2020, corresponding to 37.8~fb$^{-1}$, of which 34.6~fb$^{-1}$ were taken at a COM energy of $10.58$~GeV, corresponding to the $\Upsilon(4S)$ resonance, and 3.2~fb$^{-1}$ were taken 60~MeV below the $\Upsilon(4S)$ resonance.\par
The Belle II detector consists of a vertex detector (VXD), a central drift chamber (CDC), time of propagation counter and proximity focusing aerogel ring imaging Cherenkov detectors used for particle identification (PID), an electromagnetic calorimeter (ECL), and a $K_\textrm{L}$ and muon detector (KLM). The sub-detectors are described in detail in Ref.~\cite{b2book}.\par
\begin{figure}[h]
\centering
\includegraphics[width=0.65\textwidth]{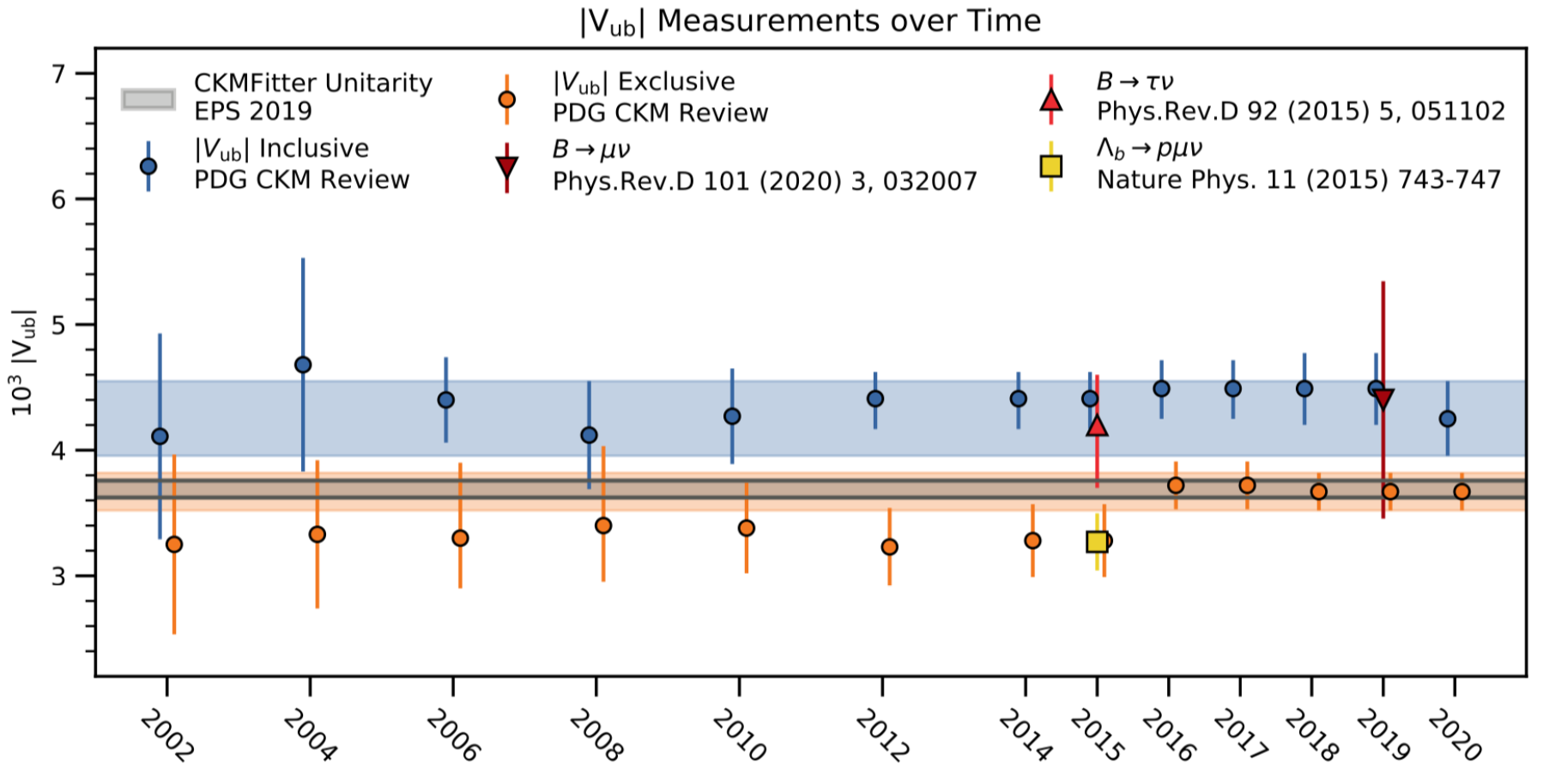}
\vspace{-0.3cm}
\caption{Measurements of $|V_{ub}|$ over time; the blue points show the results of inclusive $|V_{ub}|$ measurements \cite{pdg}, while the orange points show the results of exclusive measurements \cite{pdg}. The blue and orange bands show the combined uncertainty of inclusive and exclusive measurements, respectively. The red triangles show the results of purely leptonic searches, $B\to \tau \nu$ \cite{taunu} (red triangle) and $B\to \mu \nu$ \cite{munu} (red upside down triangle). The yellow square shows the result from $\Lambda_b \to p \mu \nu$ \cite{lambda}. The grey band shows the average obtained by the CKMFitter group \cite{ckmfitter}\cite{prim}. }
\label{fig:vub}
\end{figure}

One of the core goals of the Belle II physics program is the precision measurement of Cabbibo-Kobayashi-Maskawa (CKM) matrix elements. Semileptonic decays of $B$ mesons play a critical role in the determination of the magnitudes of the CKM quark-mixing matrix elements $|V_{cb}|$ and $|V_{ub}|$. The quark-level decay $b\to u \ell \nu$ is important for measuring the CKM matrix element $|V_{ub}|$. This is a challenging measurement due to the presence of a neutrino in the final state, which cannot be detected and manifests as missing energy in the event. This decay is suppressed compared to the decay with the charm quark in the final state, $b\to c \ell \nu$, which is the main background for this mode. Depending on the treatment of the final state meson, two different analysis approaches exist. Exclusive analyses involve explicit reconstruction of the charmless meson in specific final states. Inclusive analyses do not reconstruct the final state meson in specific final states. Previous results show tension between inclusive and exclusive measurements of the $|V_{ub}|$ matrix element \cite{pdg}. Figure~\ref{fig:vub} summarizes previous results of inclusive and exclusive $|V_{ub}|$ measurements.\par
Here we will present the first Belle II measurements of the untagged $B\to X_u e \nu_e$ decay, where $X_u$ is one or more charmless mesons, and the tagged $B^0\to \pi^- \ell^+ \nu_\ell$ exclusive decay. While the results are not yet competitive with the current world averages, they demonstrate the performance of the Belle II detector and the analysis software, and pave the way for more precise measurements in the future with larger datasets. 

\section{Untagged inclusive $B \to X_u e \nu_e$}
In this section a first search for $B \to X_u e \nu_{e}$ decays is presented. We reconstructed $B \to X_u e \nu_{e}$ decays inclusively, where only the outgoing lepton was selected. The accompanying hadron and the neutrino were not reconstructed. An untagged analysis approach was used, where the companion $B$ meson is not reconstructed. The lepton momentum endpoint was examined, to avoid the dominant background from the decay $B \to X_c e \nu_{e}$. Since the $u$ quark is lighter than the $c$ quark, the electron from $B\to X_u e \nu_e$ decay has more energy available, and thus can have a higher momentum, with electrons from $B \to X_c e \nu_e$ reaching a maximum momentum of about 2.4~GeV, and electrons from $B \to X_u \ell \nu_e$ reaching 2.8~GeV in the COM frame.\par
\begin{figure}[t!]
\centering
\begin{subfigure}[t]{0.49\textwidth}
\includegraphics[width=0.8\textwidth]{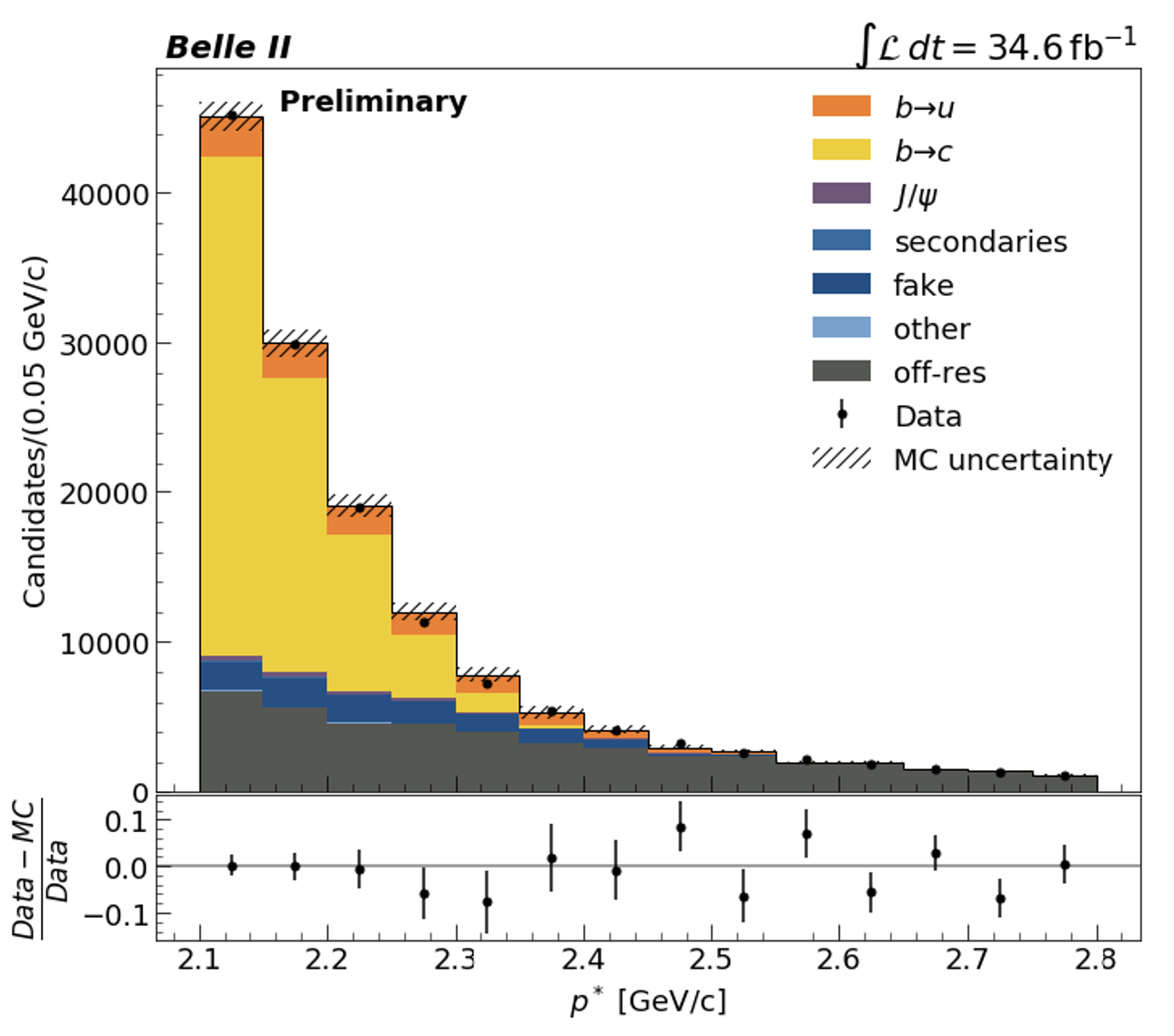}
\vspace{-0.3cm}
\caption{}
\label{data_mc_endpoint}
\end{subfigure}
\begin{subfigure}[t]{0.49\textwidth}
\includegraphics[width=0.8\textwidth]{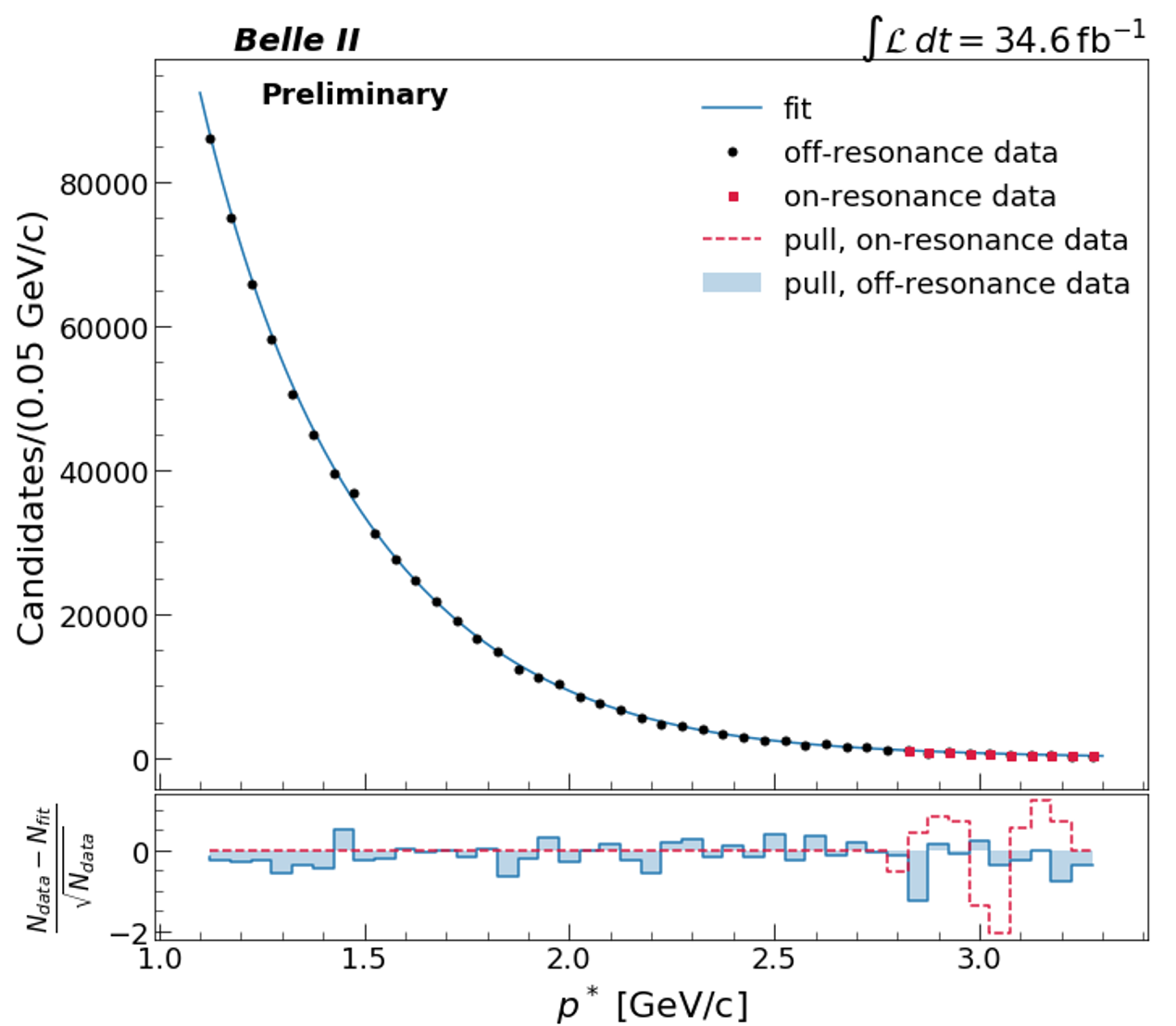}
\vspace{-0.3cm}
\caption{}
\label{offres_fit}
\end{subfigure}
\caption{(a): Comparison of data and MC electron COM momentum spectrum in the endpoint region; $b\to u$ and $b\to c$ represent all $B\to X_u e \nu_e$ and $B\to X_c e \nu_e$ events that pass the selection, respectively;  $J/\psi$ represents the electrons that originate from $J/\psi \to e^+ e^-$ decays; secondaries represent electrons that are not originating from a $B$ meson or $J/\psi$ decay; fakes represent other particles misidentified as electrons. (b): Electron spectrum in the off-resonance sample (black points) and in the on-resonance sample (red squares); the fit used to estimate the continuum contributions is shown with the blue line; the measure of goodness of fit is shown in the lower panel. }
\label{twoplots}
\end{figure}

The signal electron was identified using a selection on the electron likelihood, which combined information from several subdectors. A multivariate selection utilising event shape variables was performed to suppress the continuum backgrounds, $e^+e^- \to u\bar{u}, d\bar{d}, s\bar{s}, c\bar{c}, \ell^+\ell^-$. The most important variable in the training was the Fox-Wolfram $R_2$ moment \cite{b2book}. The signal yield was extracted from the electron momentum in the COM frame in the endpoint region, $[2.1, 2.8]$~GeV. A comparison between data and Monte Carlo (MC) of the electron spectrum at the kinematic endpoint is shown in Fig.~\ref{data_mc_endpoint}. \par
A fit was performed on the electron spectrum to estimate the remaining continuum contributions. Off-resonance data, taken 60~MeV below the $\Upsilon(4S)$ resonance, was used to estimate the continuum contributions. The fit was performed on off-resonance data in the momentum range $[1.0, 3.3]$~GeV, and simultaneously on on-resonance data in the region above 2.8~GeV, where the contributions from $B$ meson decays are not significant. A binned chi-squared minimization was used, with the following function describing the electron spectrum:
$$f(p_i^*) = a_0\cdot \exp{(a_1 p_i^* + a_2 p_i^{*^2} + a_3 p_i^{*^3})} + \exp{(a_4 p_i^* + a_5 p_i^{*^2})}, $$
where $p_i^*$ is the COM electron momentum in the $i$-th bin, $a_0, ... , a_5$ are the fit parameters. The bin width was 50 MeV. The electron spectra in the off-resonance and on-resonance data are shown in Fig.~\ref{offres_fit}. \par
\begin{figure}[h]
\centering
\includegraphics[width=0.55\textwidth]{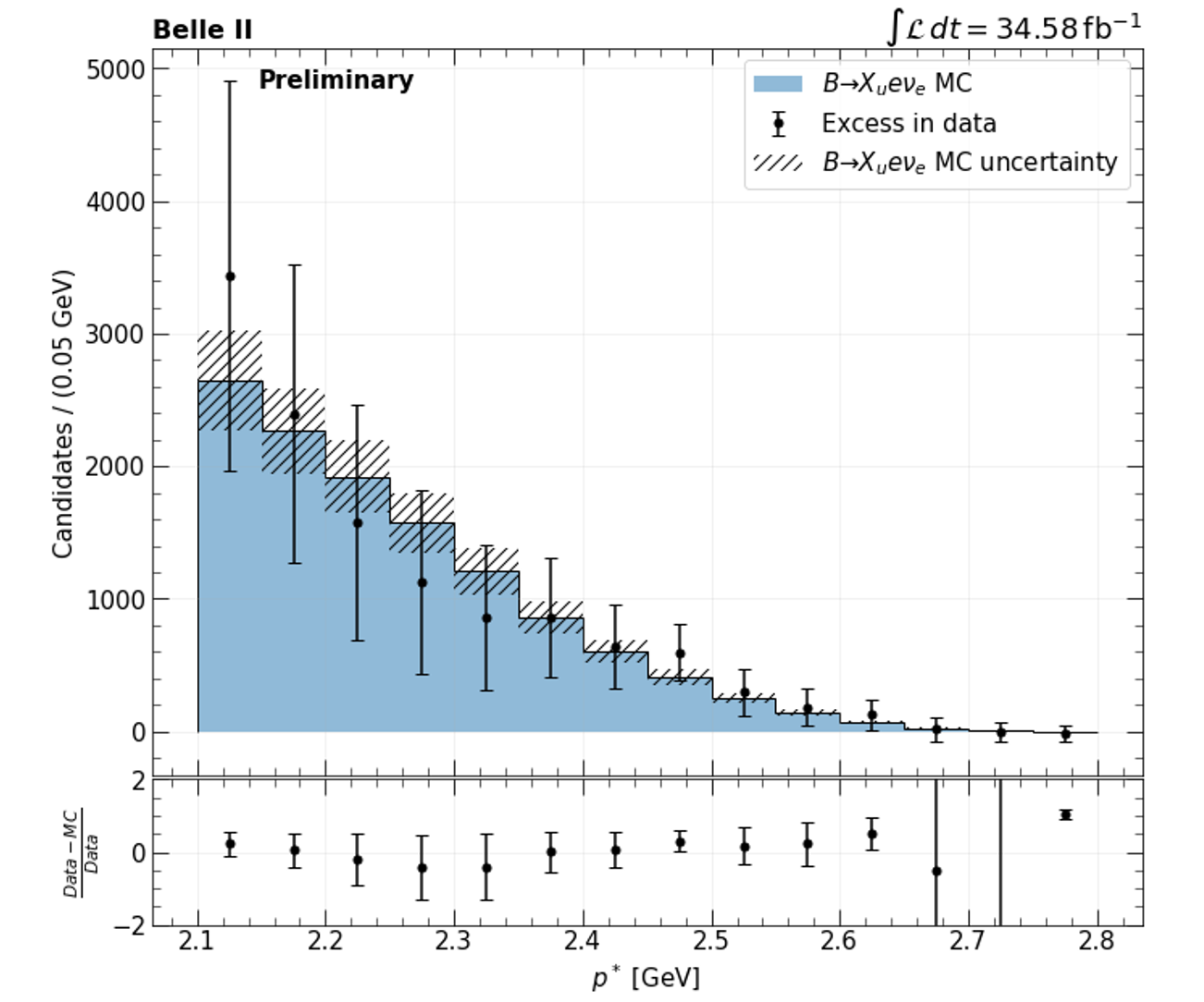}
\vspace{-0.3cm}
\caption{Comparison between $B\to X_u e \nu_e$ MC and the excess events in data in the electron endpoint region after the continuum and $B\bar{B}$ background subtraction.}
\label{excess}
\end{figure}

The backgrounds from $B\bar{B}$ decays, such as $B\to X_c e \nu$, secondary electrons not originating from $B$ meson decays, electrons from $J/\psi$ decays, and other, were estimated using a MC template fit. The binned chi-squared minimization was performed in the range $[1.0, 3.3]$~GeV in bins of 50~MeV, with the continuum backgrounds described by the results of the off-resonance fit. Continuum and background $B\bar{B}$ contributions obtained from the fits were then subtracted from the measured data in the endpoint region of the electron momentum, between 2.1 and 2.8 GeV, as shown in Fig.~\ref{excess}. The observed excess is consistent with $B \to X_u e \nu_e$ prediction, and the significance of the yield is 3$\sigma$ above zero. 

\section{Tagged exclusive $B^0\to \pi^- \ell^+ \nu_{\ell}$}
\begin{figure}[t!]
\centering
\begin{subfigure}[t]{0.3\textwidth}
\includegraphics[width=0.9\textwidth]{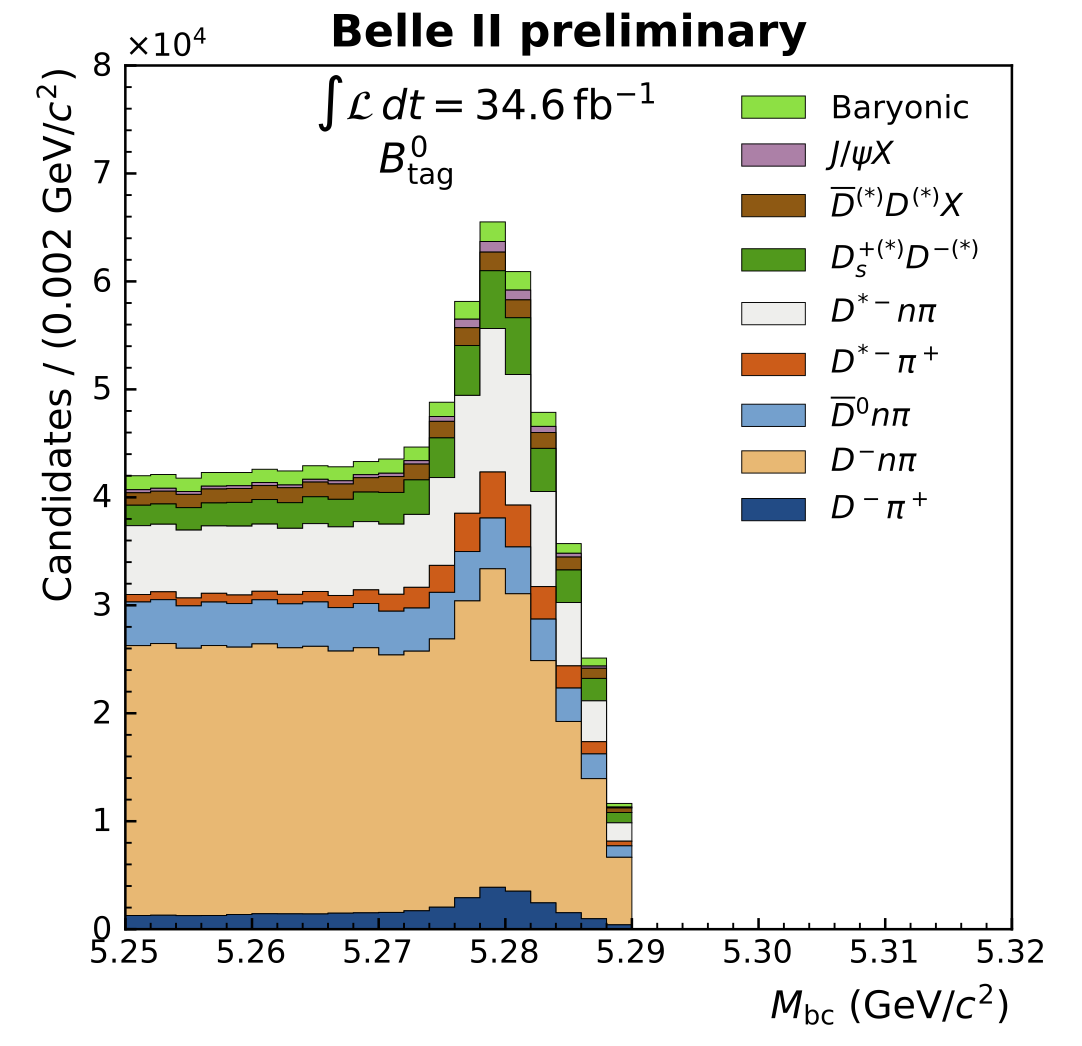}
\vspace{-0.3cm}
\caption{}
\label{hadronicb0mbc}
\end{subfigure}
~
\begin{subfigure}[t]{0.67\textwidth}
\includegraphics[width=1.0\textwidth]{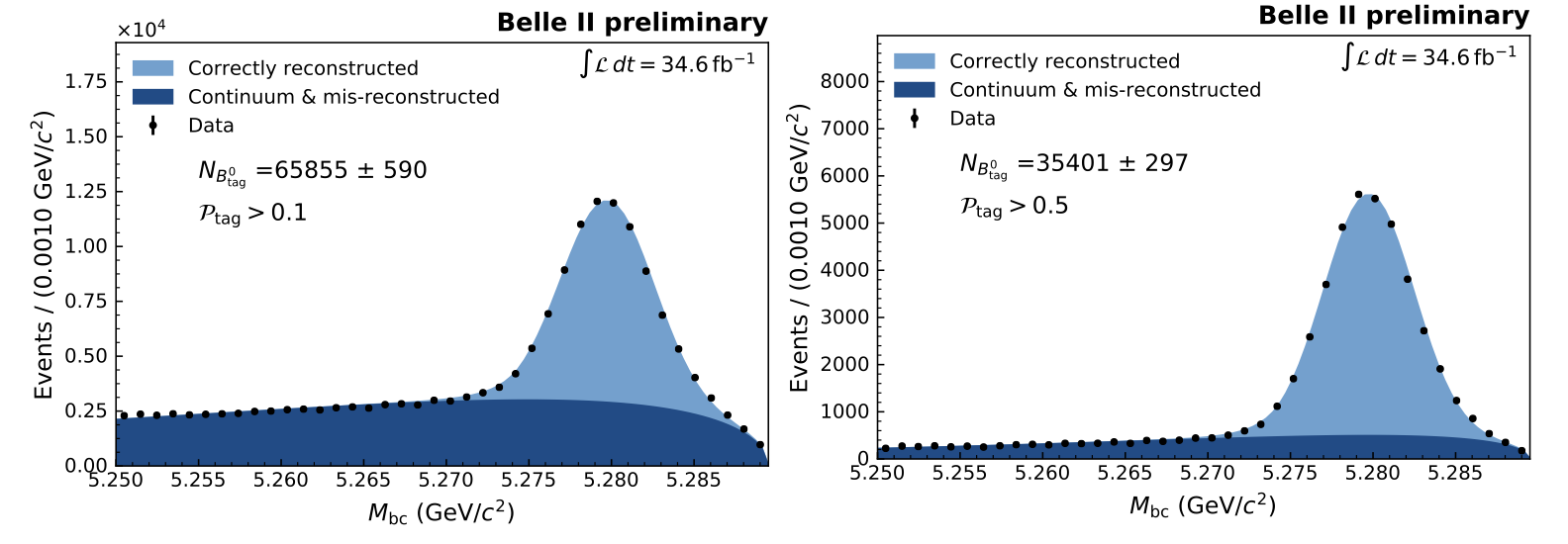}
\vspace{-0.3cm}
\caption{}
\label{feimbc}
\end{subfigure}
\caption{(a): FEI reconstructed $M_{\textrm{bc}}$  for different hadronic $B^0$ reconstruction channels; (b): FEI reconstructed $M_{\textrm{bc}}$  for different requirements on $\mathcal{P}_{\rm tag}$. Equivalent plots for $B^+$ reconstruction channels can be found in Ref.~\cite{fei2020}}
\label{feitwo}
\end{figure}

A search for the decay\footnote{Charge conjugation is implied throughout this report.} $B^0 \to \pi^- \ell^+ \nu_{\ell}$ \cite{pilnu}, where $\ell = e, \mu$, was performed using hadronic tagging provided by the Full Event Interpretation (FEI) algorithm~\cite{fei}. The FEI algorithm is a tagging algorithm developed at Belle II that uses machine learning for decay reconstruction. In FEI, one of the two $B$ mesons produced in the event is reconstructed exclusively in  one of the $\mathcal{O}(10,000)$ available hadronic or semileptonic decay modes. The algorithm uses a hierarchical reconstruction approach: first the detected tracks, clusters and vertices are used as final state particle candidates; these final state candidates are then combined to reconstruct intermediate state candidates; finally, all the candidates are combined to reconstruct the $B$ meson candidates, $B_{\rm tag}$. A dedicated multivariate classifier is trained to classify each unique 
decay within the $B$ meson decay chain.
The assigned classifier value for $B$ mesons, $\mathcal{P}_{\rm tag}$, distinguishes poorly reconstructed or misreconstructed from correctly reconstructed $B_{\rm tag}$ candidates. 
The FEI outputs multiple $B_{\rm tag}$ candidates per event, which can be further reduced by setting requirements on beam-constrained mass, $M_{\textrm{bc}}$, and energy difference with respect to the $e^+e^-$ system energy in the COM frame, $\Delta E$:
$$ M_{\textrm{bc}} = \sqrt{\frac{E^2_{\textrm{beam}}}{4} - \vec{p}_{B_{\textrm{tag}}}^2}, \quad \Delta E = E_{B_{\textrm{tag}}} - \frac{E_{\textrm{beam}}}{2},$$ 
where $ \vec{p}_{B_{\textrm{tag}}}$ and $E_{B_{\textrm{tag}}}$ are the FEI reconstructed $B_{\textrm{tag}}$ momentum and energy in the COM frame, respectively, and $E_{\rm beam}$ is the COM energy of the $e^+e^-$ system. During the hadronic FEI reconstruction, the requirements are $M_{\textrm{bc}} > $5.24~GeV and $|\Delta E| <$0.2~GeV. The distribution of $B^0_{\rm tag}$ $M_{\textrm{bc}}$  for different hadronic $B^0$ reconstruction channels is shown in Fig.~\ref{hadronicb0mbc}, and in Fig.~\ref{feimbc} for different requirements on $\mathcal{P}_{\textrm{tag}}$. \par
By exclusively reconstructing $B_{\rm tag}$ using hadronic modes, one can infer the momentum 
and direction of the signal $B$ candidate. In the COM frame the 4-momentum of the signal $B$ meson can be defined as: 
$$p_{B_\textrm{sig}} \equiv ( E_{B_{\textrm{sig}}}, \vec{p}_{B_{\textrm{sig}}} ) = \left( \frac{m(\Upsilon(4S) )}{2}, -\vec{p}_{B_{\textrm{tag}}} \right).$$
The ability to infer the four-momentum of the signal $B$ candidate makes this approach well suited for signal decays with neutrinos, which contain missing energy. \par
\begin{figure}[h]
\centering
\includegraphics[width=0.95\textwidth]{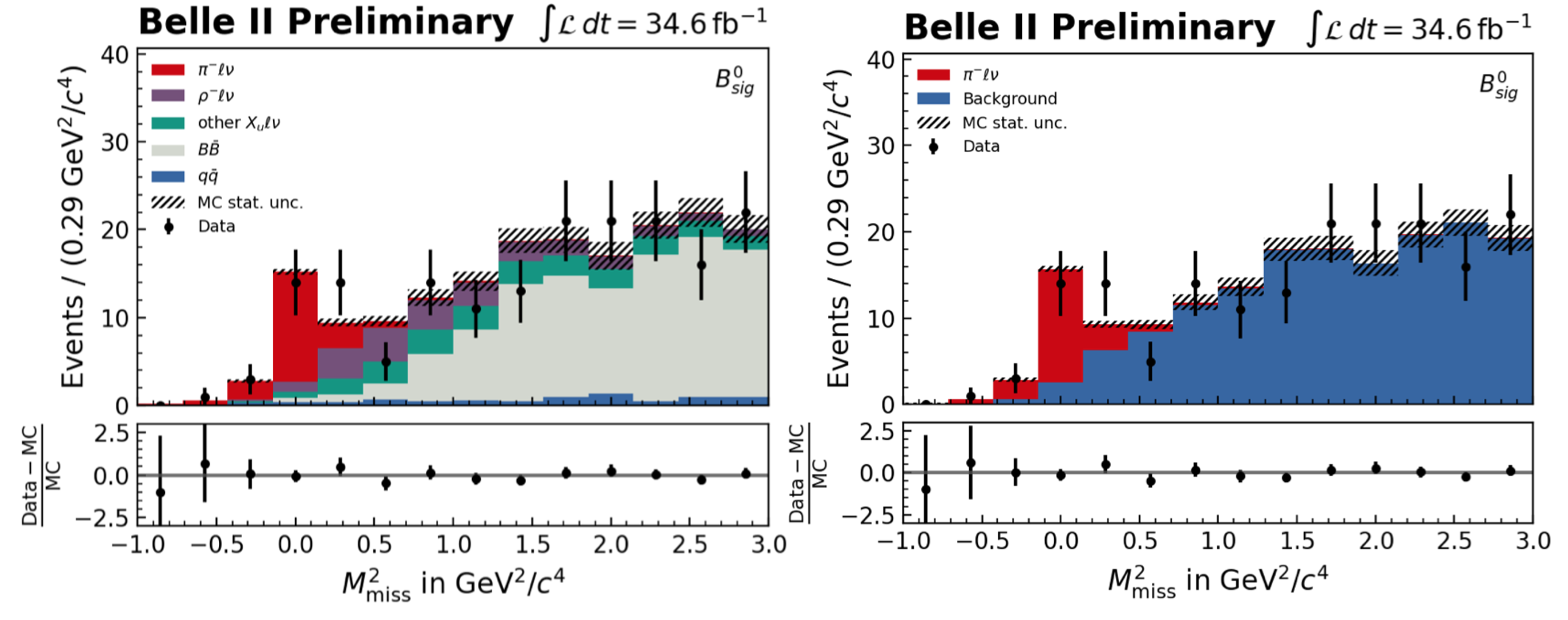}
\vspace{-0.3cm}
\caption{Distribution of $M^2_{\rm miss}$ in data and MC (left) and fitted $M^2_{\rm miss}$ distribution (right).}
\label{mmiss}
\end{figure}

In the  $B^0 \to \pi^- \ell^+ \nu_{\ell}$ analysis FEI hadronic tagging was used to identify $B_{\rm tag}$ candidates. A loose requirement on FEI classifier output, $\mathcal{P}_{\rm tag} > 0.001$, was used to maximize the signal efficiency. The requirement on $B_{\rm tag}$ $M_{\rm bc}$ was tightened to be greater than $5.27~\mathrm{GeV}$ to reduce misreconstructed candidates. A muon or electron candidate and an oppositely charged pion candidate were identified using PID likelihood variables, from the remaining tracks in the event not associated with the $B_{\rm tag}$. The continuum backgrounds were suppressed using a cut $R_2<0.4$. The signal yield was extracted using the distribution of the square of the missing mass:
$$ M^2_{\rm miss} \equiv p^2_{\rm miss} = (p_{B_{\rm sig}}-p_Y)^2,$$
where $p_Y$ is the combined four-momentum of the pion and the charged lepton.
One $\Upsilon(4S)$ candidate with the lowest $M^2_{\rm miss}$ value was retained per event. The analysis was performed blinded in the signal region $M^2_{\rm miss} < 1$ GeV$^2$. 
Fits were performed on the $M^2_{\rm miss}$ distribution to obtain the signal yields. Template probability density functions (PDFs) were constructed from signal and background $M^2_{\rm miss}$ distributions in MC. Subsequently, an extended unbinned maximum likelihood fit was performed on $M^2_{\rm miss}$ distribution in data using the template MC PDFs. 
The distribution of $M^2_{\rm miss}$ in data and MC before the fit, and fitted $M^2_{\rm miss}$ distribution are shown in Fig.~\ref{mmiss}. The resulting signal yield from the fit is $20.79 \pm 5.68$, which is in good agreement with the predicted value of 19.83 from signal MC. The observed statistical significance is 5.69$\sigma$.\par

The $B^0\to \pi^- \ell^+ \nu_\ell$ branching fraction was calculated using the following formula:
$$ \mathcal{B}(B^0 \to \pi^- \ell^+ \nu_\ell) = \frac{N_{\textrm{sig}}^{\textrm{data}}(1 + f_{+0})} {4 \cdot \textrm{C}_{\textrm{FEI}} \cdot N_{B\bar{B}} \cdot \epsilon} ,$$
where $N_{\textrm{sig}}^{\textrm{data}}$ is the fitted signal yield obtained from data, $f_{+0}$ is the ratio between the branching fractions of the decays of the $\Upsilon(4S)$ meson to pairs of charged and neutral $B$ mesons~\cite{pdg}, C$_{\textrm{FEI}}$ is the FEI data-MC calibration factor, $N_{B\bar{B}}$  is the number of $B$ meson pairs counted in the dataset, and $\epsilon$ is the reconstruction efficiency. The value of $C_{\rm FEI}$ was obtained by comparing the efficiency of FEI in data and MC using inclusive $B \to X \ell \nu$ events. It is equal to $0.83 \pm 0.03$ \cite{fei2020}. The signal efficiency determined from signal MC is (0.216 $\pm$ 0.001)$\%$. The measured $N_{B\bar{B}}$ in the dataset is $(37.711 \pm 0.602) \times 10^6$. The resulting branching fraction is $\mathcal{B} = (1.58 \pm 0.43_{\textrm{stat}} \pm 0.07_{\textrm{sys}}) \cdot 10^{-4}$. The result is in agreement with the world average value, $(1.50 \pm 0.06) \cdot 10^{-4}$~\cite{pdg}. Several sources of systematic uncertainty were identified, with the most significant being the FEI data-MC calibration. The evaluation of systematic uncertainties is described in Ref.~\cite{pilnu}.

\section{Summary and prospects for $b \to u$ measurements at Belle II}
We have presented the first Belle II measurements of $B \to X_u e \nu_e$ and $B^0 \to \pi^- \ell^+ \nu$ semileptonic $B$-meson decay modes. The measured $B^0 \to \pi^- \ell^+ \nu$ branching fraction of $(1.58 \pm 0.43_{\textrm{stat}} \pm 0.07_{\textrm{sys}})\cdot 10^{-4}$ is in agreement with previous results but statistically limited. Significant improvement on these and other $b\to u$ semileptonic measurements is expected with higher luminosities and improved understanding of the detector performance. Overall, the $|V_{ub}|$ precision will be improved to 3\% (1-2\%) inclusive (exclusive) with the full 50~ab$^{-1}$ dataset at Belle II. With improved theoretical uncertainties available, $B \to \pi \ell \nu$ is expected to be the golden mode for measurements of $|V_{ub}|$.

\end{document}